\documentstyle[aps,pra,floats,epsfig]{revtex}

\newcommand{\be}{\begin{equation}}
\newcommand{\ee}{\end{equation}}
\newcommand{\bea}{\begin{eqnarray}}
\newcommand{\eea}{\end{eqnarray}}
\newcommand{\bt}{\begin{tabular}}
\newcommand{\et}{\end{tabular}}
\newcommand{\ba}{\begin{array}}
\newcommand{\ea}{\end{array}}

\newcommand{\wt}{\widetilde}

\newcommand{\dy}{\displaystyle}

\begin{document}

\twocolumn[\hsize\textwidth\columnwidth\hsize\csname
@twocolumnfalse\endcsname

\small{\hfill{
\begin{tabular}{l}
DSF$-$17/2002
\\
quant-ph/0209018
\end{tabular}}}


\title{Multibarrier tunneling}
\author{S. Esposito$^\ast$}

\address{Dipartimento di Scienze Fisiche, Universit\`{a} di Napoli
``Federico II''\\ and \\ Istituto Nazionale di Fisica Nucleare, Sezione di
Napoli \\ Complesso Universitario di Monte S. Angelo,  Via Cinthia, I-80126
Napoli, Italy
\\ E-mail: Salvatore.Esposito@na.infn.it }

\maketitle

\begin{abstract}
We study the tunneling through an arbitrary number of finite rectangular
opaque barriers and generalize earlier results by showing that the total
tunneling phase time depends neither on the barrier thickness nor on the
inter-barrier separation. We also predict two novel peculiar features of
the system considered, namely the independence of the transit time (for non
resonant tunneling) and the resonant frequency on the number of barriers
crossed, which can be directly tested in photonic experiments. A thorough
analysis of the role played by inter-barrier multiple reflections and a
physical interpretation of the results obtained is reported, showing that
multibarrier tunneling is a highly non-local phenomenon.
\end{abstract}


\vskip2pc]


\section{Introduction}

\noindent
A renewed interest in a typical quantum phenomenon such as the tunnel
effect has been recently achieved due to a long series of experiments aimed
to measure the tunneling transit time (for reviews see, for istance,
\cite{revs}). While such experiments involving electrons are usually
difficult to realize (mainly due to the smallness of the electron de
Broglie wavelength at usual temperatures) and even of uncertain
interpretation, the observations on photonic tunneling
\cite{expt1}-\cite{expt5} has by now provided clear data on this subject.
Despite the different phenomena studied in several experiments (undersized
waveguides, photonic band gap, total internal reflection) and the different
frequency ranges for the ligth used (from the optical to the microwave
region), all such experiments have shown that, in the limit of opaque
barriers, the transit time to travel across a barrier of width $a$ is
usually {\it shorter} than the corresponding one required for real (not
evanescent) propagation through the same region of width $a$. This result
can be interpreted \cite{revs} in terms of a superluminal group velocity
$v_{\mathrm gr} >c$ which, however, does not violate Einstein causality,
since the signal velocity relevant for that \cite{somm} is never measured.
Nevertheless we prefer to look at the experimental result as an observation
of the simple Hartman effect \cite{hartman}: for opaque barriers the
tunneling phase time is independent of the barrier width. Although several
definitions of the tunneling time (also related to the different
experimental setups used) exist \cite{revs} and a general consensus on this
is still lacking, it seems that all the experimental results can be
successfully interpreted in terms of phase time \cite{univ}.
\\
Further light has been put on the problem by recent experiments involving
double barrier penetration \cite{double}. In fact, while the above effect
has been confirmed in such a system too (far from the resonances of the
structure), observations show that the transit time is also independent of
the separation distance between the barriers (supposed to be thick). This
peculair phenomenon has been studied theoretically in Ref. \cite{ORS},
where the authors have provided a straigthforward generalization of the
Hartman effect for double barrier tunneling.
\\
Convinving qualitative explanations of these two findings (namely that the
tunneling phase time is independent of the barrier thickness as well as of
the inter-barrier separation for opaque barriers) have been reported. When
considering a given wavepacket entering into a potential barrier region, a
reshaping phenomenon occurs in which the travelling edge of the pulse is
preferentially attenuated that the leading one, thus simulating a group
velocity greater than $c$ \cite{revs},\cite{expt2}. In practice the Hartman
effect in the tunneling through a thick barrier is explained from the fact
that under the barrier no phase accumulates, and the entire phase shift
comes only from the boundaries, thus being substantially independent of the
thickness \cite{superosc}.  Furthermore, when two barriers are present, the
transit time independence on the barrier separation can, instead, be
understood in terms of an effective acceleration of the forward travelling
waves in the inter-barrier region, which arises from a destructive
interference between the two barriers \cite{ORS}.
\\
Further noticeable results have been recently achieved in Ref.
\cite{superosc}, where it has been shown that a wavepacket travels {\it in
zero time} a region with $N$ arbitrary $\delta-$function barriers.
\\
In this paper we extend all these findings by considering the case of $N$
successive opaque barriers with finite widths and heigths. While we confirm
all previous results, we generalize them by showing that some peculiar
tunneling properties are independent of the number of the barriers crossed
(Sect. \ref{s2}). Furthermore, in order to establish a quantitative
interpretation of the involved phenomena, in Sect. \ref{s3} we study the
role of multiple reflections in double barrier tunneling and show how
strongly the total tunneling phase time depends on non-local effects.
Finally in Sect. \ref{s4} we discuss the results obtained and give our
conclusions.
\\
In view of the formal analogu \cite{analogy} between the Schr\"{o}dinger
equation and the electromagnetic Helmholtz equation, our study applies to
matter particle tunneling as well as to evanescent propagation of photonic
wavepackets. This is a straigthforward cosequence of the fact that in both
cases the starting point is basically the same (in our case it is Eq.
(\ref{2.2})) \cite{univ}, on interchanging the roles of angular frequency
$\omega$ and wavevector $k$ into the corresponding ones of energy $E$ and
momentum $p$ through the Planck - de Broglie relations. Thus, throughout
this paper, we use indifferently particle or wave terminology unless the
meaning of what we are doing becomes unclear.

\section{Tunneling through $N$ successive barriers}
\label{s2}

\begin{figure}[t]
\epsfysize=2.5cm
\epsfxsize=8cm
\centerline{\epsffile{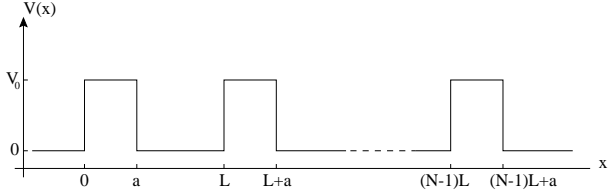} \vspace{0.8truecm}}
\caption{Potential barrier $V(x)$ with $N$ equally spaced
rectangular bumps of given heigth and width.}
\label{barrier}
\end{figure}
\noindent
Let us consider a wavepacket moving along the $x-$axis and entering at
$x=0$ into a region with a potential barrier $V(x)$ as depicted in Fig.
\ref{barrier}:
\begin{equation}\label{2.1}
V(x) \; = \; \left\{ \ba{llc} V_0 &~& (i-1)L \leq x
\leq (i-1)L+a \\ &~& \\ 0 &~& {\mathrm otherwise} \ea \right.
\end{equation}
( i=1,2,\dots,N). For the sake of simplicity we choose the heigth $V_0$ of
the potential barriers, as well as the width $a$ of each barrier, to be the
same for all $N$ rectangular barriers. We further assume equally spaced
barriers, $L-a$ being the inter-barrier distance.
\\
The propagation of the wavepacket through the barriers is described by a
scalar field $\psi$ representing the Schr\"{o}dinger wavefunction in the
particle case or some scalar component of the electric or magnetic field in
the photonic case. This is the solution of the Schr\"{o}dinger equation or the
Helmholtz equation with potential or refractive index in Eq.(\ref{2.1})
and, in both cases, it takes the following form \footnote{Obviously, the
physical field is represented by a wavepacket with a given spectrum in
$\omega$:
\[
\Psi(x,t) \; = \; \int \! d \omega \, \eta(\omega) \, \psi(x) \,
 e^{-i \omega t} ~~~,
\]
where $\eta(\omega)$ is the envelope function. Keeping this in mind, for
the sake of simplicity we deal with only stationary solutions as in Eq.
(\ref{2.2}).}:
\begin{equation}\label{2.2}
\!\!\!
\psi(x) \; = \; \left\{
\ba{llc}
\psi_{2i}(x) &~& (i-1)L \leq x \leq (i-1)L+a \\
&~& \left( i=1,2,\dots,N \right) \\ &~& \\
\psi_{2i+1}(x) &~& {\mathrm otherwise} \\
&~& \left( i=0,1,2,\dots,N \right)
\ea \right.
\end{equation}
with:
\bea
\psi_1(x) &=& e^{ikx} + R e^{-ikx} \label{2.3a} \\ &~& \nonumber \\
\psi_{2i}(x) &=& A_{2i} e^{\chi [x-(i-1)L]} + B_{2i} e^{-\chi [x-(i-1)L]} \nonumber
\\ &~& \left( i=1,2,\dots,N \right) \label{2.3b} \\ &~& \nonumber \\
\psi_{2i+1}(x) &=& A_{2i+1} e^{ik[x-(i-1)L]} + B_{2i+1} e^{-ik[x-(i-1)L]} \nonumber
\\ &~& \left( i=1,2,\dots,N-1 \right) \label{2.3c} \\ &~& \nonumber \\
\psi_{2N+1}(x) &=& T e^{ik[x-(N-1)L]}  ~~~. \label{2.3d}
\eea
As noted in Ref. \cite{univ}, the explicit dependence on the frequency of
the (real) wavevector $k$ in the barrier-free regions and imaginary
wavevector $i \chi$ in the barrier ones enters only in the final expression
for the phase time. As long as possible we do not use a particular
dispersion relation in order to draw general features which are common to
the particle and to the wave case.
\\
The $4N$ unknown coefficients $R,T,A_{i},B_{i}$ are obtained from the $4N$
matching conditions for the function $\psi$ and its derivative
$\psi^\prime$ at the discontinuity points $x+(i-1)L$, $x=(i-1)L+a$ of the
potential:
\begin{equation}\label{2.4}
\ba{rcl}
\psi_{2i-1} \left( x=(i-1)L \right) &=& \psi_{2i} \left( x=(i-1)L \right)
\\ &~& \\
\psi_{2i-1}^\prime \left( x=(i-1)L \right) &=& \psi_{2i}^\prime \left( x=(i-1)L \right)
\\ &~& \\
\psi_{2i} \left( x=(i-1)L+a \right) &=& \psi_{2i+1} \left( x=(i-1)L+a \right)
\\ &~& \\
\psi_{2i}^\prime \left( x=(i-1)L+a \right) &=& \psi_{2i+1}^\prime \left( x=(i-1)L+a \right)
\ea
\end{equation}
($i=1,2,\dots,N$). Note that the quantities $R$ and $T$ have the meaning of
(total) reflection and transmission coefficient from the $N-$barrier
system, respectively, and satisfy the unitarity condition:
\begin{equation}\label{2.5}
|R|^2 + |T|^2 \; = \; 1 ~~~.
\end{equation}
We have produced a {\texttt Mathematica} symbolic code in order to obtain
explicit analytic expressions for all the coefficients appearing in Eqs.
(\ref{2.3a})-(\ref{2.3d}). However, we here report only the interesting
result obtained for the transmission coefficient $T(N)$ for an $N-$barrier
system in the opaque barrier approximation $\chi a >>1$. In this limit the
quantity $T(N)$ can be factorized in the following way:
\begin{equation}\label{2.6}
T(N) \, e^{ika} \; = \; {\cal C}_0 \, {\cdot} \, {\cal E}(N) \, {\cdot} \, {\cal F}(N)
\end{equation}
\bea
   {\cal C}_0 &=& \frac{4 i \chi k}{(k+i \chi)^2}  \nonumber
\\ {\cal E}(N) &=& \left[ e^{- \chi a} \right]^N \nonumber
\\ {\cal F}(N) &=& \left[ \frac{2 \chi k}{2 \chi k \cos k(L-a) -
(k^2-\chi^2) \sin k(L-a)} \right]^{N-1} \nonumber
\eea
Note that only the real terms ${\cal E}$ and ${\cal F}$ depend on $a,L,N$,
while the complex factor ${\cal C}_0$ does not. As a consequence, since the
tunneling phase time $\tau$ is defined as:
\begin{equation}\label{2.7}
\tau \; = \; \frac{d\phi}{d\omega}
\end{equation}
and the quantity
\bea
\phi & \equiv & \arg \left\{ T(N) \, e^{ika} \right\} \; = \; \arg \left\{
\frac{4 i \chi k}{(k+i \chi)^2} \right\} \; = \nonumber
\\ &=& \arctan \, \frac{k^2-\chi^2}{2 \chi k} \label{2.8}
\eea
is independent of $a,L,N$, we arrive at the general conclusion that {\it
the tunneling phase time for a system of $N$ opaque barriers depends
neither on the barrier width and inter-barrier distance nor on the number
of the barriers}.
\\
Let us now discuss the effects of the real terms in Eq.(\ref{2.6}) on the
tunneling probability $P_T(N) = |T(N)|^2$:
\bea
&~& P_T(N) \; = \; \left[ \frac{4 \chi k}{k^2 + \chi^2} \right]^2
\left[ e^{- \chi a} \right]^{2N} \, {\cdot} \nonumber \\
&~&
\left[ \frac{2 \chi k}{2 \chi k \cos k(L-a) \, - \,
(k^2-\chi^2) \sin k(L-a)} \right]^{2(N-1)} \label{2.9}
\eea
We easily recognize that the last factor in Eq.(\ref{2.9}), coming from the
term ${\cal F}(N)$ is responsible of the resonance structure of the
transmission probability. The factor ${\cal F}(N)$ is, of course, absent in
the case of only one barrier, i.e. $N=1$ or $N \neq 1$ but $L=a$. However,
no resonance can occur even in the particular case in which the
inter-barrier distance is tuned in a way that:
\begin{equation}\label{2.10}
L-a \; = \; \frac{\nu \pi}{k} ~~~~~ \left( \nu =0,1,2,\dots \right) ~~~.
\end{equation}
In this case, waves moving forward and backward in the inter-barrier
regions interfere between them such that no resonance takes place.
\\
The resonance condition for the tunneling probability is, from Eq.
(\ref{2.9}), the following:
\begin{equation}\label{2.11}
\tan k(L-a) \; = \; \frac{2 \chi k}{k^2 - \chi^2} ~~~.
\end{equation}
It is wortwhile to observe that Eq.(\ref{2.11}) does not depend on $N$, so
that {\it the resonant frequency is the same irrespective of the number of
barriers to be crossed}.
\\
Finally, we point out an intriguing consequence of the resonance condition.
Let us write Eq.(\ref{2.11}) as follows:
\begin{equation}\label{2.12}
\tan \phi \ {\cdot} \ \tan k (L-a) \; = \; 1
\end{equation}
where $\phi$ is given in Eq.(\ref{2.8}), and take the derivative of Eq.
(\ref{2.12}) with respect to the angular frequency $\omega$. By using Eq.
(\ref{2.7}) we easily arrive at the following relation:
\begin{equation}\label{2.13}
\tau \ + \ \tau_0 \; = \; 0 ~~~,
\end{equation}
where $\tau_0$ is the (phase) time for travelling the inter-barrier
distance $L-a$ in vacuum. Keeping in mind that the total tunneling time has
the same value of the tunneling time for crossing only one barrier (see
above), we see that, when resonant tunneling takes place, {\it the total
time required to cover the distance $L$} (one barrier length $a$ plus one
inter-barrier distance $L-a$) {\it is zero}.

\section{Multiple reflections and non-locality}
\label{s3}

\noindent
In order to have a physical interpretation of the results obtained
previously, we now consider the effect of single barriers on the
propagation of the wavepacket through the entire $N-$barrier system, by
invoking the superposition principle. For the sake of simplicity, we will
study the case of a system of two opaque barriers.

\subsection{Partial coefficients}

\noindent
For $N=2$, in the barrier-free regions, Eqs.(\ref{2.3a})-(\ref{2.3d})
reduce to the following:
\bea
\psi_1(x) &=& e^{ikx} + R e^{-ikx}  \nonumber \\
\psi_3(x) &=& A_3 e^{ikx} + B_3 e^{-ikx}  \label{3.1} \\
\psi_5(x) &=& T e^{ik(x-L)}  \nonumber
\eea
where the explicit expressions for the coefficients are reported in
Appendix \ref{app}. Let us now denote with $R_1,T_1$ and $R_2,T_2$ the
(partial) reflection and transmission coefficients of the first and second
barrier, respectively. In the region with $x<0$ the reflected wave is
described by the term:
\begin{equation}\label{3.2}
R e^{-ikx} \; = \; R_1 e^{-ikx} + B_3 T_1 e^{-ikx} ~~~,
\end{equation}
while for $x>L+a$ the transmitted one is described by:
\begin{equation}\label{3.3}
T e^{ik(x-L)} \; = \; A_3 T_2 e^{ik(x-L)} ~~~.
\end{equation}
By taking into account multiple reflections from the two barriers in the
region with $a<x<L$, we see that the forward travelling wave is described
by the term:
\begin{equation}\label{3.4}
A_3 e^{ikx} \; = \; T_1 \left[ 1 + R_1 R_2 + \left( R_1 R_2 \right)^2 +
\dots \right] e^{ikx} ~~~,
\end{equation}
while the backward one is described by:
\begin{equation}\label{3.5}
B_3 e^{-ikx} \; = \; A_3 R_2 e^{-ik(x-L)} ~~~.
\end{equation}
Then, by introducing the quantity:
\begin{equation}\label{3.6}
S \; = \; \sum_{l=0}^{\infty} \left( R_1 R_2 \right)^l \; = \;
\frac{1}{1-R_1 R_2} ~~~,
\end{equation}
which accounts for multiple reflections, from Eqs.(\ref{3.2})-(\ref{3.5})
we obtain:
\begin{equation}\label{3.7}
\ba{rcl}
R &=& R_1 + B_3 T_1 \\ T &=& A_3 T_2 \\ A_3 &=& T_1 S
\\ B_3 &=& A_3 R_2 e^{ikL} ~~~.
\ea
\end{equation}
By solving these equation with respect to the partial reflection and
transmission coefficients, we get:
\begin{equation}\label{3.8}
\ba{rcl}
R_1 &=& \dy \frac{R-A_3 B_3}{1-B_3^2} \\ & & \\ T_1 &=& \dy \frac{A_3 - B_3
R}{1-B_3^2} \\ & & \\ R_2 &=& \dy \frac{B_3}{A_3} e^{-ikL} \\ & & \\ T_2
&=& \dy \frac{T}{A_3} ~~~.
\ea
\end{equation}
In the opaque barrier limit $\chi a >>1$, for the second barrier we obtain:
\begin{equation}\label{3.9}
\ba{rcl}
R_2 &=& R_{\mathrm OB} \ e^{ikL} \\ T_2 &=& T_{\mathrm OB} \ e^{ikL} ~~~,
\ea
\end{equation}
while for the first barrier:
\begin{equation}\label{3.10}
\ba{rcl}
R_1 &=& R_{\mathrm OB} + R_{\mathrm Q} + R_{\mathrm R} \\ T_1 &=&
T_{\mathrm OB} + T_{\mathrm Q} + T_{\mathrm R} ~~~,
\ea
\end{equation}
where:
\begin{equation}\label{3.11}
\ba{rcl}
R_{\mathrm OB} &=& \dy \frac{k-i\chi}{k+i\chi} \left[ 1 - \frac{4i\chi
k}{(k+i\chi)^2} e^{-2 \chi a} \right] \\ & &
\\ T_{\mathrm OB} &=& \dy \frac{4i\chi k}{(k+i\chi)^2} e^{-ika} e^{-\chi a}
\ea
\end{equation}
are the reflection and transmission coefficients corresponding to a
one-barrier system ($N=1$) and:
\begin{equation}\label{3.12}
\ba{rcl}
   R_{\mathrm Q} &=& \dy - \left( \frac{k-i\chi}{k+i\chi} \right)^3 {\cal F}^2
e^{2ik(L-a)} e^{-2 \chi a} \\ & &
\\ R_{\mathrm R} &=& \dy \left( \frac{k-i\chi}{k+i\chi} \right)^3 {\cal F}^2
e^{ikL} e^{-2 \chi a} \\ & &
\\ T_{\mathrm Q} &=& \dy \left( \frac{k-i\chi}{k+i\chi} \right)^2 {\cal F}
e^{2ik(L-a)} e^{-ikL} e^{-\chi a} \\ & &
\\ T_{\mathrm R} &=& \dy - \left( \frac{k-i\chi}{k+i\chi} \right)^2 {\cal F}
e^{-\chi a} ~~~.
\ea
\end{equation}
For future reference, we also consider the partial coefficients
$R_1^0,T_1^0,R_2^0,T_2^0$ in the approximation of no multiple reflections,
as considered in Ref. \cite{ORS} \footnote{The authors of Ref. \cite{ORS}
have considered the case of no multiple reflections and, moreover, they
also neglected the second term $B_3 T_1$ in the first equation in
(\ref{3.7}) corresponding to backward waves in the $x<0$ region transmitted
from the first barrier, reflected from the second one and again transmitted
from the first barrier. While their parametrization of the wavefunction is,
of course, permitted and leads to correct results, nevertheless the partial
coefficients they obtained have no direct physical meaning, as we will show
below.}. These are obtained from Eqs.(\ref{3.7}) by setting $S=1$. We have
\footnote{In the approximation considered in Ref. \cite{ORS} the quantity
$R_1^0$ in Eq.(\ref{3.13}) should be replaced by the following one:
\[
\wt{R}_1^0 \; = \; R_{\mathrm OB} + \frac{k-i\chi}{k+i\chi} {\cdot}
\frac{4i\chi k}{(k+i\chi)^2} {\cdot} {\cal F} e^{ik(L-a)} e^{-2\chi a} ~~~.
\]}:
\begin{equation}\label{3.13}
\ba{rcl}
R_1 &=& R_{\mathrm OB} + R_{\mathrm Q} \\ T_1 &=& T_{\mathrm OB} +
T_{\mathrm Q} ~~~,
\ea
\end{equation}
while $R_2^0,T_2^0$ are the same as in Eqs.(\ref{3.9}).

\subsection{Unitarity conditions}

\noindent
The interpretation of the quantities $R_1,T_1$ and $R_2,T_2$ as reflection
and transmission coefficients of the first and second barrier is derived
from the unitarity conditions satisfied by these coefficients. In fact,
since $|R|^2+|T|^2=1$ and $|R_{\mathrm OB}|^2+|T_{\mathrm OB}|^2=1$, we
find that:
\begin{equation}\label{3.14}
\ba{rcl}
|R_1|^2+|T_1|^2 &=& 1 \\ |R_2|^2+|T_2|^2 &=& 1 ~~~.
\ea
\end{equation}
It is easily recognizable as well that, assuming no multiple reflection,
the total probability for scattering from the first barrier is {\it lower}
than 1 \footnote{Instead, by using the parametrization of Ref. \cite{ORS},
we obtain an unphysical scattering probability {\it greater} than 1,
\[
|\wt{R}_1^0|^2+|T_1^0|^2 \; = \; 1 + {\cal F}^2 e^{-2 \chi a} ~~~,
\]
which makes impossible to give a direct physical meaning to
$\wt{R}_1^0,T_1^0$.}:
\begin{equation}\label{3.15}
|R_1^0|^2+|T_1^0|^2 \; = \; 1 - {\cal F}^2 e^{-2 \chi a} ~~~,
\end{equation}
this revealing that something has been forgotten. Obviously, multiple
reflections are the missing term and it is worth to observe that the
probability for this phenomenon to occur, which from Eq.(\ref{3.15}) we
deduce to be ${\cal F}^2 e^{-2 \chi a}$, is given by:
\begin{equation}\label{3.16}
P_{\mathrm R} \; \equiv \; |R_{\mathrm R}|^2+|T_{\mathrm R}|^2 \; = \;
{\cal F}^2 e^{-2 \chi a} ~~~.
\end{equation}
Thus the quantities $R_{\mathrm R}$ and $T_{\mathrm R}$, that must be added
to the no multiple reflection coefficients $R_1^0$ and $T_1^0$ in order to
obtain the complete ones $R_1$ and $T_1$ respectively, can be interpreted
as the terms describing the phenomenon of multiple reflections between the
first and second barrier.
\\
The meaning of the picture just outlined is then, quite trivial. $R_2$ and
$T_2$ corresponding to the second barrier are simply given by the
one-barrier coefficients $R_{\mathrm OB},T_{\mathrm OB}$ times a phase
factor which takes into account the fact that this barrier starts at $x=L$,
while the reference point in our discussion is at $x=0$. Instead, $R_1$ and
$T_1$ corresponding to the first barrier are given by the sum of two terms:
the first one is the no multiple reflection coefficients while the second
one describes the phenomenon of multiple reflections. However, it is
remarkable that {\it no multiple reflection coefficients $R_1^0$ and
$T_1^0$ in Eqs.(\ref{3.13}) do not coincide with the one-barrier
coefficients $R_{\mathrm OB}$ and $T_{\mathrm OB}$}. This is an obvious
consequence of the fact that the scattering probability from the first
barrier, {\it neglecting} multiple reflections, cannot be equal to the
unity and the extra terms $R_{\mathrm Q}$ and $T_{\mathrm Q}$ in Eqs.
(\ref{3.13}) must be present in order to achieve the probability constraint
in Eq.(\ref{3.15}). On the other hand, the scattering probability, {\it
including} multiple reflections, must be equal to 1 (according to Eq.
(\ref{3.14})), so that we can deduce that the quantities $R_{\mathrm Q}$
and $T_{\mathrm Q}$ are related to the multiple reflection coefficients
$R_{\mathrm R}$ and $T_{\mathrm R}$. It is very easy to obtain from Eqs.
(\ref{3.12}) that $R_{\mathrm Q}$ and $T_{\mathrm Q}$ differ from
$R_{\mathrm R}$ and $T_{\mathrm R}$ just for a phase factor (depending on
$L$ and $a$):
\begin{equation}\label{3.17}
\frac{R_q}{R_{\mathrm R}} \; = \; \frac{T_{\mathrm Q}}{T_{\mathrm R}} \; = \; - e^{ik(L-2a)} ~~~.
\end{equation}
Then, multibarrier tunneling is a highly non-local phenomenon driven by
multiple reflections, whose influence on the determination of the
reflection and transmission coefficients is (indirectly) present even in
the case in which they are neglected.

\subsection{Tunneling phase time}
\label{s3c}

\noindent
Let us now consider the tunneling phase time $\tau$ in Eq.(\ref{2.7})
corresponding to the  double barrier crossing here considered and introduce
the quantities:
\bea
\phi_1 &=& \arg \left\{ T_1 \ e^{ika} \right\} \nonumber \\
\phi_2 &=& \arg \left\{ T_2 \ e^{ika} \right\} \label{3.18} \\
\phi_{\mathrm S} &=& \arg \left\{ S \ e^{ik(L-a)} \right\} \nonumber
\eea
whose derivatives with respect to frequency give the phase times for the
first barrier crossing, the second barrier crossing and the time associated
to multiple reflections, respectively. Since $T=T_1 T_2 S$ from Eqs.
(\ref{3.7}), the total tunneling phase is given by:
\begin{equation}\label{3.19}
\phi \; = \; \phi_1 + \left( \phi_2 - kL \right) + \phi_{\mathrm S} ~~~.
\end{equation}
This relation leads to the obvious conclusion that the tunneling time
$\tau$ is the sum of the partial times $\tau_1$ and $\tau_2$
\footnote{Note that the time $\tau_2$ corresponds to the phase
$\phi_2 - kl$, since the travelling along the distance $L$ is already taken
into account in $\phi_1 + \phi_{\mathrm S}$ or, in other words, in the
expression for the coefficient $T_2$ in (\ref{3.9}) we have already
considered the shift from $x=0$ to $x=L$.} spent to travel across the first
and second barrier, respectively, plus the time $\tau_{\mathrm S}$ required
by multiple reflections in the inter-barrier region of length $L-a$.
However, it is interesting to evaluate the explicit expressions for these
times and, from Eqs.(\ref{3.18}) we get:
\bea
\phi_1 &=& \phi_0 - \frac{kL}{2} + ka  \label{3.20} \\
\phi_2 - kL &=& \phi_0  \label{3.21}
\eea
where $\phi_0 = \arg \{ T_{\mathrm OB} e^{ika} \}$ is the one-barrier
tunneling phase time and \footnote{For opaque barriers, the leading term in
$S$ is, from Eqs.(\ref{3.6}),(\ref{3.9}),(\ref{3.10}):
\[
S \; = \; \frac{(k+i\chi)^2}{4i\chi k} e^{-ikL/2} \frac{2 \chi k}{2 \chi k
\cos kL/2 - (k^2 - \chi^2) \sin kL/2} ~~~.
\]
}
\begin{equation}\label{3.22}
\phi_{\mathrm S} \; = \; - \phi_0 + \frac{kL}{2} - ka ~~~.
\end{equation}
While the time to cross the second barrier equals exactly the one-barrier
tunneling phase time (see Eq.(\ref{3.21})), from Eqs.(\ref{3.20}) and
(\ref{3.22}) we see that:
\begin{equation}\label{3.23}
\phi_1 + \phi_{\mathrm S} \; = \; 0 ~~~,
\end{equation}
that is {\it the time spent to travel from the starting edge of the first
barrier to the starting edge of the second one is zero}. Something similar
to this statement has already been suggested in literature (see, for
istance, Ref. \cite{ORS}), but now we have a quantitative proof for that.
Moreover, we can also deduce that, due to multiple reflections, the time to
cross the first barrier is usually {\it different} from the one-barrier
tunneling phase time since:
\begin{equation}\label{3.24}
\phi_0 - \phi_1 \; = \; \frac{\phi_{\mathrm Q} - \phi_{\mathrm R}}{2}
\end{equation}
where
\bea
\phi_{\mathrm Q} &=& \arg \left\{ T_Q \right\} \; = \; 2 \phi_0 + kL - 2 ka
\nonumber \\
\phi_{\mathrm R} &=& \arg \left\{ T_R \right\} \; = \; 2 \phi_0 \nonumber
\eea
are the phase times corresponding to the terms $T_{\mathrm Q}$ and
$T_{\mathrm R}$, the equality holding true only in the case in which the
inter-barrier distance coincide with the barrier width, i.e. $L=2a$.

\section{Conclusions}
\label{s4}

\noindent
In this paper we have studied the tunneling of a particle or a photonic
wavepacket through an arbitrary number $N$ of finite rectangular opaque
barriers and obtained an analytic expression for the total transmission
coefficient Eq.(\ref{2.6}). From this we have confirmed and generalized to
the present case what was found earlier for a system of one \cite{hartman}
or two \cite{ORS} barriers: the (total) tunneling phase time is independent
both of the barrier width and of inter-barrier distance. These features
have been observed experimentally for single \cite{expt1}-\cite{expt5} and
double barrier \cite{double} tunneling using photonic setups.
\\
Amazingly enough, we have further found that, although the tunneling
probability decreases exponentially with the barrier thickness and with the
number of barriers (in the opaque barrier limit), the tunneling time does
not depend even on the number of barriers crossed, i.e. it is the same for
one, two or more barriers. Moreover, when considering resonant tunneling,
we have also shown that the position in frequency (or energy) of the
resonance of the structure is independent of the number of barriers as
well. These two novel predictions can be experimentally tested using,
again, photonic devices.
\\
In order to obtain a physical picture of what happens in the system
considered and, especially, of the peculiar features outlined above, we
have studied the role of multiple reflections between the barriers on the
tunneling and found this to be a highly non-local phenomenon. In fact, as
shown in Sect. \ref{s3}, even in the case of increasingly large separation
between the barriers, the effect of multiple reflections cannot be avoided
at all. In particular multiple reflections play a crucial role in the
understanding of the intriguing results on the (total) tunneling time
quoted above. Though in Sect. \ref{s3} we have dealt with a two-barrier
system, the main results achieved can be easily generalized to multibarrier
tunneling as follows. For $N$ barriers the partial reflection and
transmission coefficients corresponding to the first $N-1$ barriers are
clearly influenced by multiple reflections occurring in the inter-barrier
regions, while those associated to the last barrier are not and coincide
with one-barrier coefficients up to a phase factor. In particular, as shown
in Sect. \ref{s3c}, the tunneling phase time for crossing only the last
barrier equals that for a single barrier structure. Since the total
tunneling time for crossing all the barriers coincides as well with the
one-barrier time (see Sect. \ref{s2}), we immediately deduce that the time
for travelling from the starting edge of the first barrier to the starting
edge of the last one is zero. Note that such a result can be achieved only
if we taken into account multiple reflections and, in any case, the partial
times for crossing single barriers are usually different from the
one-barrier tunneling time.
\\
Finally we point out that our findings also agree with the recent results
reported in \cite{superosc}, according to which a wavepacket travels in
zero time a region with $N$ $\delta-$function barriers. In fact, as said
above, the total tunneling time coincide with the transit time for the last
barrier or one-barrier phase time. From \cite{univ} (see Eq.(13) of that
paper) we then see that, for $\chi \rightarrow \infty$, this time tends to
zero, thus recovering the result of Ref. \cite{superosc}. It would then be
nice, in the future, to make the connection between multiple reflections
studied here and the tunneling interpretation in terms of superoscillations
quoted in \cite{superosc}.

\acknowledgements
The appearance of this paper is entirely due to the kind encouragement of
Prof. E. Recami. Many useful discussions with him and with Dr. G. Salesi
and O. Pisanti have been greatly appreciated.

\appendix

\section{Coefficients for $N=2$}
\label{app}

\noindent
From Eqs.(\ref{2.4}) we obtain the following expressions for the
coefficients describing the propagation through two successive opaque
barriers:
\bea
R & \simeq & \frac{k-i\chi}{k+i\chi} \left[ 1 + 2i \sin k(L-a) \ {\cal F} \
e^{-2\chi a}\right]
\\
A_2 & \simeq & \frac{2k}{k-i\chi} \ \frac{(k-i\chi)^2}{2\chi k} \ \sin
k(L-a)  \ {\cal F} \ e^{-2\chi a}
\\
B_2 & \simeq & \frac{k-i\chi}{k+i\chi} \left\{ \frac{2k}{k-i\chi} \left[ 1
- \frac{(k-i\chi)^2}{2\chi k} \ \sin k(L-a) \ {\cdot} \right. \right. \nonumber \\
&~& \left. \left. {\cdot} \ {\cal F} \ e^{-2\chi a} \right] \right\}
\\
A_3 & \simeq & e^{-ikL} \ {\cal F} \ e^{-\chi a}
\\
B_3 & \simeq & \frac{k-i\chi}{k+i\chi} \ e^{ikL} \ {\cal F} \ e^{-\chi a}
\\
A_4 & \simeq & 0
\\
B_4 & \simeq & \frac{2k}{k+i\chi} \ {\cal F} \ e^{- \chi a}
\\
T & \simeq & \frac{4i\chi k}{(k+i\chi)^2} \ {\cal F} \ e^{-2\chi a}
\eea
(in all these expressions we have neglected terms of third order in
$e^{\chi a}$).

\end{document}